\documentclass[a4paper,final]{appolb}
\pdfoutput=1
\usepackage[pdftex]{graphicx}
\usepackage[unicode=true, bookmarks=false,bookmarksnumbered=true, bookmarksopen=true, breaklinks=false, backref=false, pagebackref=false, colorlinks=true]{hyperref}
\hypersetup{pdftitle={Landau Gauge Fixing on GPUs}, pdfauthor={Nuno Cardoso, Pedro Bicudo, Paulo J. Silva and Orlando Oliveira}, linkcolor=black, citecolor=black, urlcolor=black, filecolor=black, pdfpagelayout=OneColumn, pdfnewwindow=true, pdfstartview=XYZ, plainpages=false}
\usepackage{color}
\usepackage{eurosym}
\usepackage{bm}
\usepackage{amsmath}
\usepackage{amssymb}
\usepackage{amsmath, amsfonts, epsfig, xspace}
\usepackage{array}
\usepackage{tabularx}
\usepackage{booktabs}
\usepackage{enumerate}
\usepackage{algorithm,algorithmic}
\usepackage{pgf}
\usepackage{verbatim}
\usepackage{braket}
\usepackage[T1]{fontenc}
\usepackage[english]{babel}
\usepackage{subfig}
\usepackage{float}
\usepackage{url}
\usepackage{listings}
\usepackage{algorithm}
\usepackage{algorithmic}

\newcommand\T{\rule{0pt}{2.0ex}}
\newcommand\B{\rule[-0.6ex]{0pt}{0pt}}

\usepackage{verbatim}

\begin{document}

\title{Landau Gauge Fixing on GPUs and String Tension
\thanks{Presented by N. Cardoso at the International Meeting "Excited QCD", Peniche, Portugal, 6 - 12 May, 2012}}
\author{Nuno Cardoso, Pedro Bicudo
\address{CFTP, Departamento de F\'{i}sica, Instituto Superior T\'{e}cnico, Universidade T\'{e}cnica de Lisboa, Av. Rovisco Pais, 1049-001 Lisbon, Portugal\newline}
\\
{Paulo J. Silva and Orlando Oliveira
}
\address{CFC, Departamento de F\'{\i}sica, Faculdade de Ci\^encias e Tecnologia, Universidade de Coimbra, 3004-516 Coimbra, Portugal}
}

\maketitle

\begin{abstract}
We explore the performance of CUDA in performing Landau gauge fixing in Lattice QCD,
using the steepest descent method with Fourier acceleration.
The code performance was tested in a Tesla C2070, Fermi architecture.
We also present a study of the string tension at finite temperature in the confined phase.
The string tension is extracted from the color averaged free energy and from the color singlet using Landau gauge fixing.
\end{abstract}

\section{Introduction}

The string tension $\sigma(T)$ is a relevant order parameter to study confinement. 
While above the deconfinement temperature $T_c$ the simplest order parameter is the Polyakov loop, below $T_c$ the expectation value of a single Polyakov loop vanishes.  Below $T_c$, to study the decrease of confinement with $T$ a new order parameter must be used, and we use the string tension. 

\begin{figure}[!htb]
\begin{centering}
    \subfloat[\label{fig:string_uncorrected}]{
\begin{centering}
    \includegraphics[width=6.0cm]{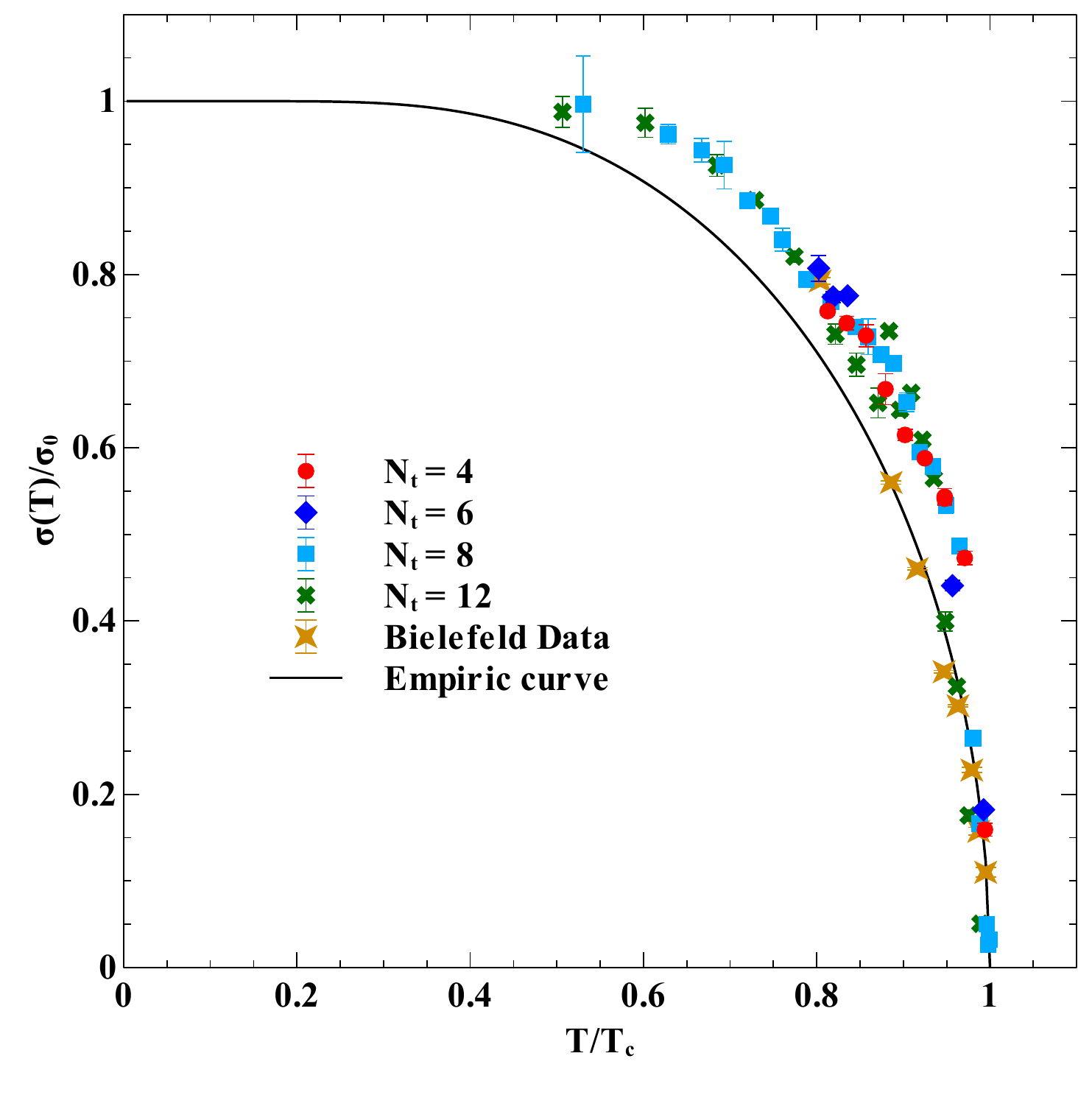}
\par\end{centering}}
    \subfloat[\label{fig:string_corrected}]{
\begin{centering}
    \includegraphics[width=6.0cm]{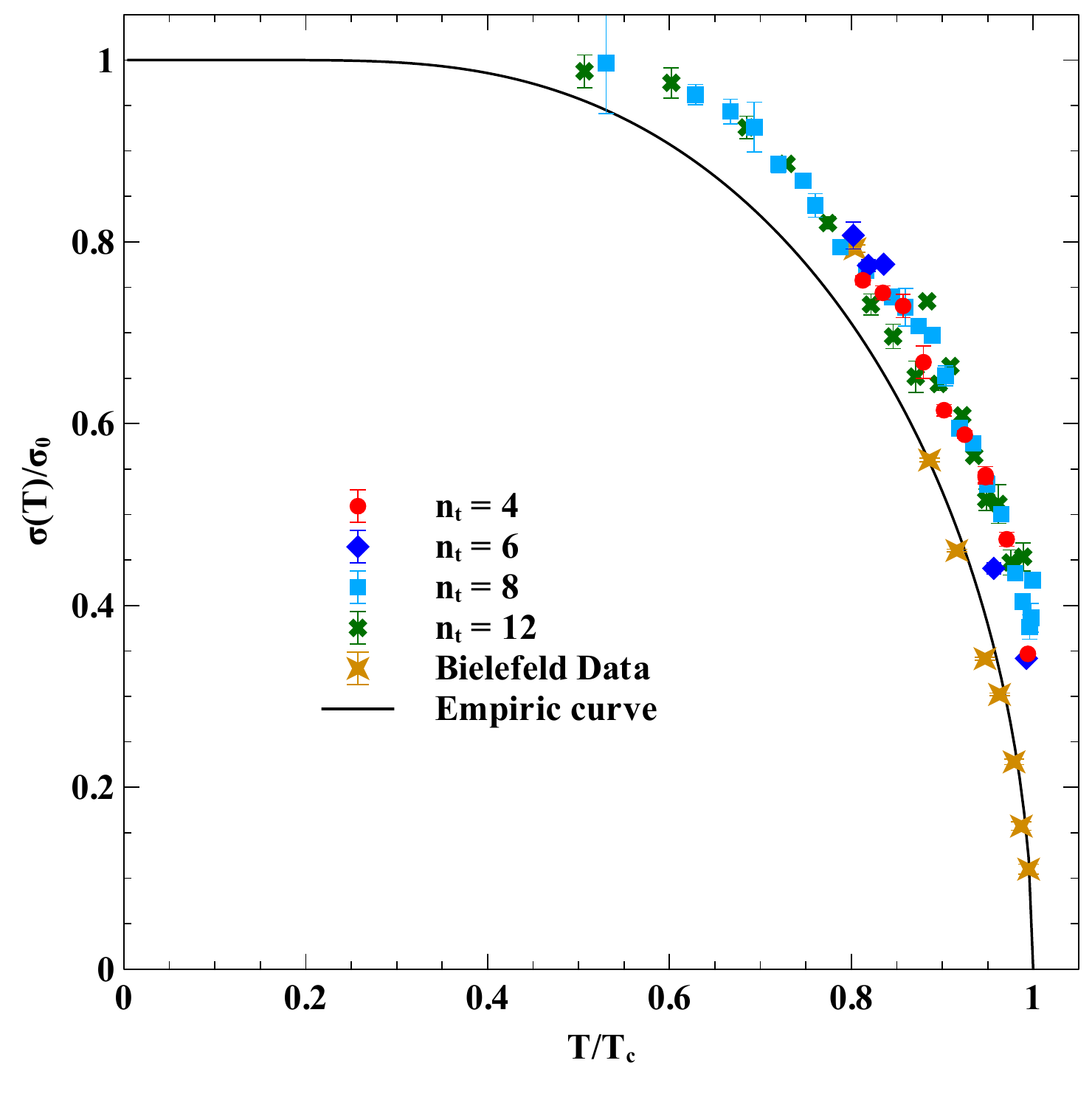}
\par\end{centering}}
\par\end{centering}
    \caption{String tension, $\sigma/\sigma_0$, as function of the temperature for lattices $48^3\times N_t = 4, 6, 8, 12$, see \cite{Cardoso:2011hh}, combined with the results from the Bielefeld group, \cite{Kaczmarek:1999mm} . The black line corresponds to the ferromagnet magnetization $M/M_{sat}$ critical curve, \cite{Bicudo:2010hg}. \protect\subref{fig:string_uncorrected} string tension with contaminated configurations near the phase transition and \protect\subref{fig:string_corrected} string tension with removed contaminated configurations.}
    \label{fig:stringtension}
\end{figure}

The existing lattice QCD results for the string tension critical curve have been computed by the Bielefeld group \cite{Kaczmarek:1999mm} and Cardoso et al. \cite{Cardoso:2011hh}, Fig. \ref{fig:stringtension}, who have studied in detail the heavy quark potential, from the color average free energy, at finite temperature in the confined phase. 
Their results confirmed a first order deconfinement phase transition, as expected for SU(3).
Bicudo \cite{Bicudo:2010hg} compared the string tension points obtained by Bielefeld group \cite{Kaczmarek:1999mm}
with different order parameter curves and found empirically that the ferromagnetic critical curve is the one closer to the Bielefeld data, Fig. \ref{fig:stringtension}.

In Fig. \ref{fig:stringtension}, we show the results for the string tension at finite temperature, for more details see \cite{Cardoso:2011hh}.
Near the phase transition, it is necessary to analyse the histogram of the Polyakov loop history, if a double peak structure is found in the histogram then there are configurations in the wrong phase. 
In Fig. \ref{fig:string_uncorrected} and Fig. \ref{fig:string_corrected}, we show the results with and without those contaminated configurations, respectively.

In this paper we will study the string tension extracted from the singlet energy,
\begin{equation}
	e^{-F_\text{singlet}(r,T)/T+C}=\frac{1}{3}\Braket{\Tr \left(W(\vec{r})\, W^\dagger(0)\right)}\, ,
\end{equation}
where $W(.)$ is the temporal Wilson line. However, this is not gauge invariant and therefore we must fix the gauge.
Here we will use the Landau gauge fixing using the steepest descent method with Fourier acceleration, \cite{Davies:1987vs,Oliveira:2003wa}. In order to improve the signal-to-noise-ratio, we apply the multi-hit method before the Landau gauge fixing.

In section 2 we discuss the CPU and GPU implementation of the Landau gauge fixing algorithm. In section 3, we show the benchmark results and the string tension results in the Landau gauge extracted from the singlet energy.
In section 4, we conclude.


\section{Parallel Implementation of the Landau Gauge Fixing}

\subsection{CPU Implementation}
The MPI parallel version was implemented in C++, using the machinery provided by the Chroma library \cite{Edwards2005}.
The Chroma library is built on top of QDP++, a library which provides a data-parallel programming environment suitable for Lattice QCD.
The use of QDP++ allows the same piece of code to run both in serial and parallel modes.
For the Fast Fourier transforms, the code uses PFFT, a parallel FFT library written by Michael Pippig \cite{Pippig2011}.
Note that, in order to optimize the interface with the PFFT routines, we have compiled QDP++ and Chroma using a lexicographic layout.

\subsection{GPU Implementation}

For the parallel implementation of Landau gauge fixing on GPUs \cite{Cardoso:2012pv},  we used version 4.1 of CUDA. 
For the 4D dimensional lattice, we address one thread per lattice site; using 3D thread blocks, we only need to reconstruct the other lattice index inside the kernel.
Although CUDA supports up to 3D thread blocks \cite{NVIDIA:cudac}, the same does not happen for the grid, which can be up to 2D or 3D depending on the architecture and CUDA version. Nevertheless, the code is implemented with 3D thread blocks and for the grid we adapted the code for each situation.
We use the GPU constant memory to put most of the constants needed by the GPU, like the number of points in the lattice, using \lstinline!cudaMemcpyToSymbol()!.
To store the lattice array in global memory, we use a SOA type array as described in \cite{Cardoso:2011xu}.
The main reason to do this is due to the implementation of the FFT algorithm. 
The FFT is applied for all elements of $\Delta(x)$ matrix separately. Using the SOA type array, the FFT can be applied directly to the elements without the need of copying data or data reordering.
For the Fourier transforms, the code uses CUDA CUFFT \cite{CUFFT}. Since there is no direct support for 4D FFTs, we will use 2D plus 2D FFTs with \lstinline!cufftPlanMany()!.

\begin{table}[!htb]
\begin{centering}
\begin{tabular}{|c|c|c|c|}
\hline
\multicolumn{4}{c}{\T\B \textbf{NVIDIA Tesla C2070}}\tabularnewline
\hline
\hline
\T\B Number of GPUs  & 1 & Global memory & 6 GB\tabularnewline
\hline
\T\B CUDA Capability & 2.0 & Mem. bandwidth (ECC off) & 148 GB/s\tabularnewline
\hline
\T\B Multiprocessors (MP) & 14 & Shared mem. (per SM) KB & 48 or 16\tabularnewline
\hline
\T\B Cores per MP & 32 & L1 cache (per SM) KB & 16 or 48\tabularnewline
\hline
\T\B Total \# of cores & 448 & L2 cache (chip wide) & 768KB\tabularnewline
\hline
\T\B  Clock rate & 1.15 GHz & ECC support & yes \tabularnewline
\hline
\end{tabular}
\par\end{centering}
\caption{NVIDIA's graphics card specifications used in this work.}
\label{tab:nvidia_gpu_specs}
\end{table}

In order to reduce memory traffic we can use the unitarity of SU(3) matrices and store only the first two rows (12 real numbers) and reconstruct the third row on the fly when needed, instead of storing it.

\section{Results}

\begin{figure}[!htb]
\begin{centering}
    \subfloat[\label{fig:perf_32_3_ecc_off}]{
\begin{centering}
    \includegraphics[width=11.2cm]{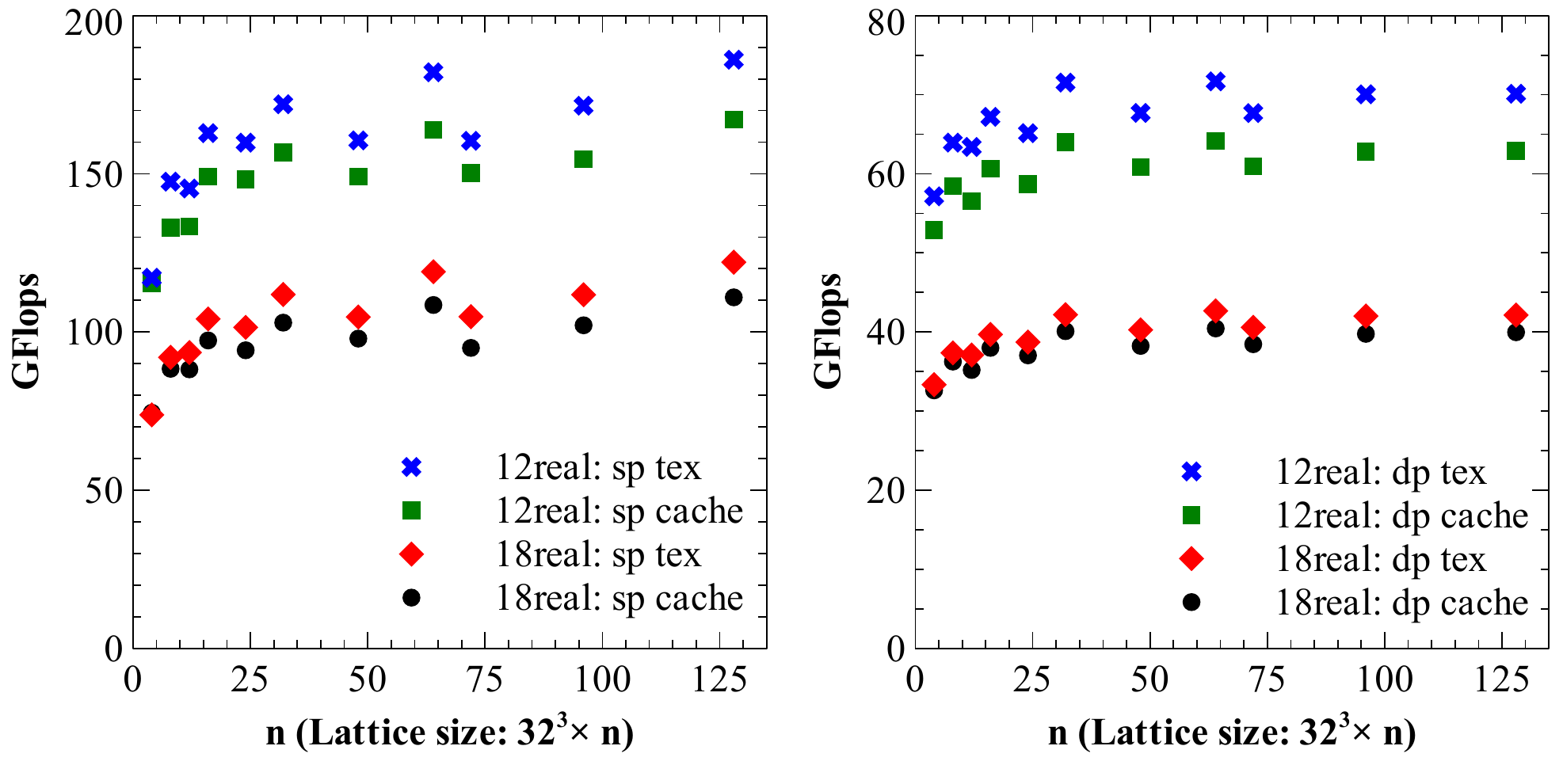}
\par\end{centering}}

    \subfloat[\label{fig:perf_32_3_ecc_on}]{
\begin{centering}
    \includegraphics[width=11.2cm]{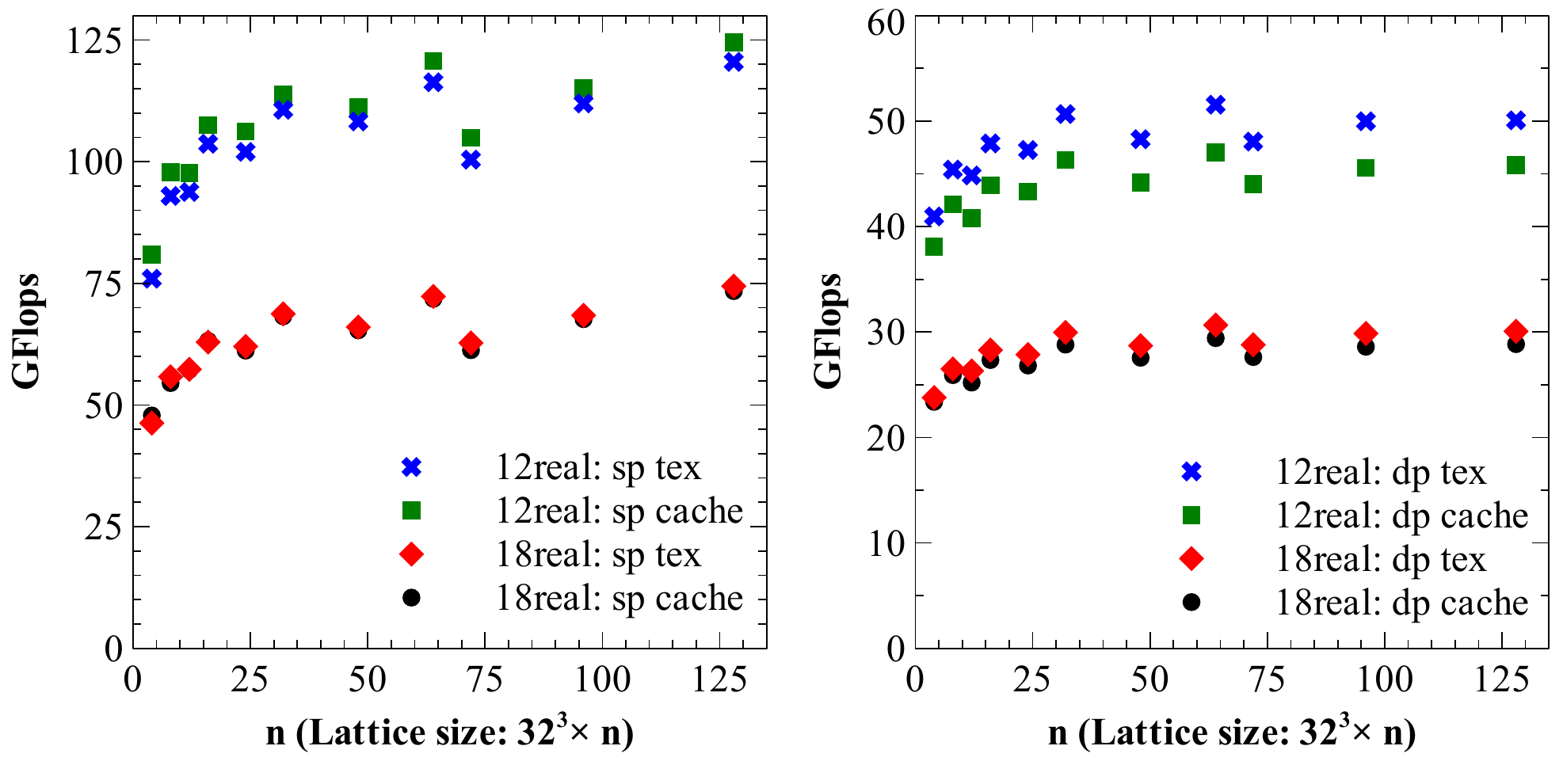}
\par\end{centering}}
\par\end{centering}
    \caption{Performance in GFlops. \protect\subref{fig:perf_32_3_ecc_off} with ECC Off and \protect\subref{fig:perf_32_3_ecc_on} with ECC On. sp: single precision, dp: double precision, 18real: full SU(3) matrix, 12real: 12 real parametrization, tex: using texture memory and cache: using L1 and L2 cache memory.}
    \label{fig:perf_32_3}
\end{figure}

In this section we show the benchmarks for the steepest descent Fourier accelerated code for Landau gauge
fixing in lattice QCD and the results for the string tension in the Landau gauge extracted from the singlet energy.

The benchmark runs using the MPI implementation were performed on the Centaurus cluster at the University of Coimbra. The Centaurus has 8 cores per node, each node has 24 GB of RAM, with 2 intel Xeon E5620@2.4 GHz (quad core) and has a DDR Infiniband interconnecting the various nodes.
The benchmark runs on GPU were performed at Instituto Superior T\'{e}cnico on a
NVIDIA Tesla C2070, Table \ref{tab:nvidia_gpu_specs}, and using version 4.1 of CUDA.

\subsection{Performance Results}
In Fig. \ref{fig:perf_32_3}, we show the GPU performance, in GFlops, of the algorithm using a 12 parameter reconstruction and the full (18 number) representation in single and double precision. 
The GPU memory access using the L1 and L2 cache and the texture memory was compared. The best performance is achieved using texture memory and the 12 real number parametrization.

\begin{figure}[!htb]
\begin{center}
    \includegraphics[width=9.4cm]{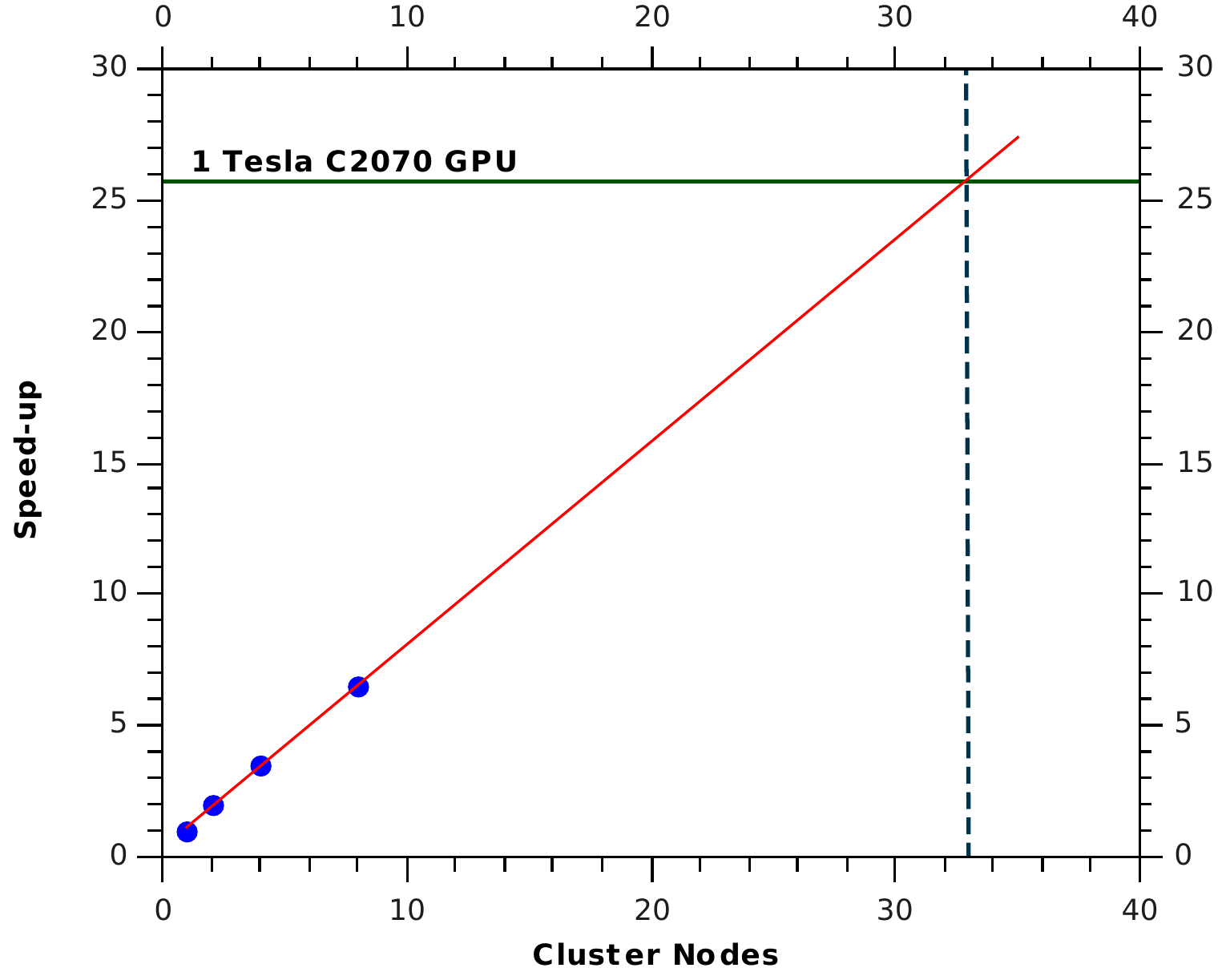}
\end{center}
\caption{Strong scaling CPU tested for a $32^4$ lattice volume and comparison with the GPU for the best 
              performance, 12 real number parametrization, ECC Off and using texture memory in double precision.
             }
    \label{fig:cpuvsgpu}
\end{figure}

The CPU MPI implementation shows a good linear scaling against the number of computing nodes, Fig. \ref{fig:cpuvsgpu}. We test the GPU (using 12 real parametrization, texture memory and ECC off),  and the CPU performance for a $32^4$ lattice volume and for $\beta = 5.8,6.0,6.2$ in double precision.
In order to have the same performance as the GPU, the CPU MPI implementation requires 32 computing codes, i.e., 256 cores.

\subsection{String Tension in the Landau gauge}

\begin{figure}[!htb]
\begin{center}
    \includegraphics[width=9.9cm]{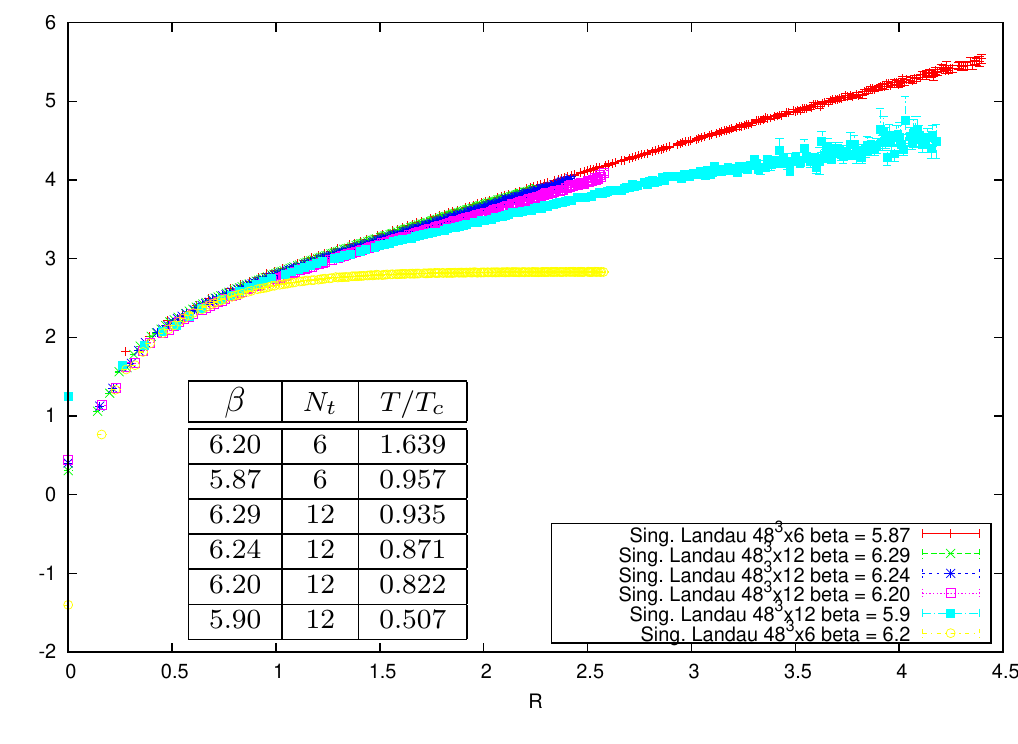}
\end{center}
\caption{Singlet energy as function of the distance in units of the zero temperature string tension.}
    \label{fig:stringsinglet}
\end{figure}

Fig. \ref{fig:stringsinglet} shows the results for the singlet energy in the Landau gauge.

The singlet energy is fitted with the potential $a_0 - a_1/r + a_2 r$, where $a_1$ is fixed to the L\"{u}scher Coulomb $\pi/12$ term and $a_2$ provides $\sigma(T)/\sigma_0$.
For $T>T_c$, the string tension is zero and in agreement with the color average free energy.
However, for $T<T_c$, the string tension, $\sigma(T)/\sigma_0$, show a temperature independent behaviour, i.e., $\sigma \sim 0.7\sigma_0$.
The string tension extracted from the color singlet in Landau gauge is constant and temperature independent in the confined phase.

\section{Conclusions}

From all the runs using a C2070 Tesla GPU, peak performance was measured as 186/71 GFlops for single/double precision.
From the performance point of view, a run on a single GPU delivers the same performance as the CPU code when running on 32 nodes (256 cores), if one assumes a linear speed-up behavior.

In the Landau gauge, the string tension, at finite temperature, extracted from the singlet energy shows a constant behaviour in the confined phase.

Our GPU code can be downloaded from the Portuguese Lattice QCD collaboration homepage \cite{ptqcd}.

\section*{Acknowledgments}

This work was partly funded by the FCT contracts,  POCI/FP/81933/2007, 
CERN/FP/83582/2008, PTDC/FIS/100968/2008, CERN/FP/109327/2009, CERN/FP/116383/2010 and CERN/FP/123612/2011.

Nuno Cardoso and Paulo Silva are supported by FCT under the contracts SFRH/BD/44416/2008 and SFRH/BPD/40998/2007 respectively.
We would like to thank NVIDIA Corporation for the hardware donation used in this work via Academic Partnership 
program.


\end{document}